\documentclass{JINST}
\usepackage{graphicx}
\usepackage{txfonts}
\usepackage{textcomp}
\usepackage{multirow}
\usepackage{natbib}
\graphicspath{{figures/}}

\title{Advanced modelling of the {\sc Planck}-LFI radiometers}

\def\instLABEN{$^{1}$}
\def\instUNIMI{$^{2}$}
\def\instIASFMI{$^{3}$}
\def\instIASFBO{$^{4}$}
\def\instIFP{$^{5}$}
\def\instJBO{$^{6}$}
\def\instOATS{$^{7}$}
\def\instNUM{$^{8}$}
\def\instIAC{$^{9}$}
\def\instDA{$^{10}$}
\def\instELEMANCH{$^{11}$}
\def\instYLI{$^{12}$}
\def\instUCSB{$^{13}$}
\def\instMILLI{$^{14}$}

\author{P.~Battaglia{\instLABEN}\thanks{Corresponding Author, e--mail: paola.battaglia@thalesaleniaspace.com} ,
      C.~Franceschet{\instUNIMI}
  ,
      A.~Zonca{\instIASFMI}
  ,
      M.~Bersanelli{\instUNIMI}
  ,
      R.C.~Butler{\instIASFBO}
  ,
      O.~D'Arcangelo{\instIFP}
  ,
      R.J.~Davis{\instJBO}
  ,
      S.~Galeotta{\instOATS}
  ,
      P.~Guzzi{\instNUM}
  ,
      R.~Hoyland{\instIAC}
  ,
      N.~Hughes{\instDA}
  ,
      P.~Jukkala{\instDA}
  ,
      D.~Kettle{\instELEMANCH}
  ,
      M.~Laaninen{\instYLI}
  ,
      R.~Leonardi{\instUCSB}
  ,
      D.~Maino{\instUNIMI}
  ,
      N.~Mandolesi{\instIASFBO}
  ,
      P.~Meinhold{\instUCSB}
  ,
      A.~Mennella{\instUNIMI}
  ,
      P.~Platania{\instIFP}
  ,
     L.~Terenzi{\instIASFBO}
  ,
     J.~Tuovinen{\instMILLI}
  ,
     J.~Varis{\instMILLI}
  ,
     F.~Villa{\instIASFBO}
  ,
     A.~Wilkinson{\instJBO}\\
\llap{\instLABEN} Thales Alenia Space Italia S.p.A.,\\
 S.S. Padana Superiore 290, 20090 Vimodrone (Mi), Italy \\
 \llap{\instUNIMI}
 Universit\'a di Milano, Dipartimento di Fisica,\\
 Via G.~Celoria 16, I-20133 Milano, Italy\\
 \llap{\instIASFMI} INAF-IASF Milano,\\
 Via E.~Bassini 15, I-20133 Milano, Italy\\
\llap{\instIASFBO}
 INAF-IASF Bologna, \\
 Via P.~Gobetti, 101, I-40129 Bologna, Italy\\
 \llap{\instIFP}
 IFP-CNR \\
 via Cozzi 53, 20125 Milano\\
 \llap{\instJBO}
 Jodrell Bank Centre for Astrophysics\\
 Alan Turing Building, The University of Manchester, Manchester, M13 9PL, UK \\
 \llap{\instOATS} INAF-OATs,\\
 Via G.B.~Tiepolo 11, I-34131, Trieste, Italy\\
 \llap{\instNUM} Numonyx, R\&D Technology Center,\\ 
 Via C.~Olivetti 2, 20041 Agrate Brianza (MI), Italy\\
 \llap{\instDA}
 DA-Design Oy\\
 Jokioinen, Finland\\
 \llap{\instELEMANCH} School of Electrical and Electronic Engineering,\\ 
 The University of Manchester, Manchester, M60 1QD, UK\\
 \llap{\instYLI}
 Ylinen Electronics Oy\\
 Kauniainen, Finland\\
 \llap{\instUCSB} Department of Physics,\\
 University of California, Santa Barbara, CA 93106-9530, USA\\ 
 \llap{\instMILLI}
 MilliLab\\
 VTT Technical Research Centre of Finland, Espoo, Finland\\
 \llap{\instIAC}
 Instituto de Astrofìsica de Canarias,\\  C/ Via Làctea S/N, E-38200, La Laguna (Tenerife), Spain
}

\abstract{
The Low Frequency Instrument (LFI) is a radiometer array covering the 30-70 GHz spectral range on-board the ESA {\sc Planck}
satellite, launched on May 14th, 2009 to observe the cosmic microwave background (CMB) with unprecedented precision. 

In this paper we describe the development and validation of a software model of the LFI pseudo-correlation receivers which
enables to reproduce and predict all the main system parameters of interest as measured at each of the 44 LFI detectors.
These include system total gain, noise temperature, band-pass response, non-linear response. The LFI Advanced RF Model
(LARFM) has been constructed by using commercial software tools and data of each radiometer component as measured at single unit level.

The LARFM has been successfully used to reproduce the LFI behavior observed during the LFI ground-test campaign.
The model is an essential element in the database of LFI data processing center and will be available for any detailed
study of radiometer behaviour during the survey.
}


\begin{document}

\section{Introduction}
The properties of the Cosmic Microwave Background (CMB) contain a wealth of information about physical conditions
in the early universe and a great deal of effort has gone into measuring those properties since its discovery.
Numerous ground based and balloon borne experiments, as well as space missions have been devised to obtain measurements
of CMB spectrum, spatial anisotropies, and polarisation over a wide range of wavelengths and angular scales. 
Following the Cosmic Background Explorer (COBE\footnote{http://lambda.gsfc.nasa.gov/product/cobe/}) and the Wilkinson 
Microwave Anisotropy Probe (WMAP\footnote{http://map.gsfc.nasa.gov/}) satellites, launched by NASA in 1989 and 2001
respectively, {{\sc Planck}\footnote{http://www.esa.int/planck}} is an ESA space mission designed to extract essentially
all information encoded in the temperature anisotropies and to measure CMB polarisation to high accuracy.

{\sc Planck}, successfully launched on 2009 May, the 14th, will observe the entire sky from a Lissajous orbit around the second Lagrange point L2
of the Earth-Sun system. The focal plane formed by the 1.5 m off-axis telescope hosts two cryogenic instruments: the
Low Frequency Instrument (LFI), covering 30-70 GHz in three bands \cite{2009_LFI_cal_M2}, and the High Frequency
Instrument (HFI) which covers the range 100 GHz to 857 GHz in six bands \cite{2009_HFI_Instrument}. The two instruments
operate at $20 K$ and at $0.1 K$ respectively. Such temperatures are achieved through a combination of passive radiative
cooling and three active coolers \cite{2009_COM_Mission}. While LFI and HFI alone have unprecedented capabilities,
it is the combination of data from the two instruments that gives {\sc Planck} the imaging power, the redundancy and the
control of systematic effects and foreground emissions needed to achieve the scientific goals of the mission.

The LFI has been designed to cover the low frequency portion of the {\sc Planck} spectral range with three bands centred at
30, 44 and 70 GHz using pseudo-correlation receivers based on Indium Phosphide cryogenic high electron mobility transistors
(HEMT) low noise amplifiers (LNA). 

In order to support the radiometers design and testing phase as well
as the study the instrument behaviour and potential
systematic effects during the survey we have constructed a software model
of the receivers, the LFI Advanced RF Model (LARFM).
In this paper we describe the development and validation of the LARFM,
which has been constructed by using commercial
software tools. The model was developed in two stages: the first
stage, the Analytical LARFM (section~\ref{sec:analytical}), is an analytical,
parametric model built upon the requirements and specifications of the
LFI receivers frequency by frequency. Its development started before
the hardware was built and proved its usefulness in predicting
radiometers performance and to provide first-order estimates of the
expected impact of potential systematics.

The second stage is the Real-Data LARFM (section~\ref{sec:realdata}),
implemented after the LFI was built, which includes the data of each
radiometer component as measured at single unit level channel by
channel. The model is capable of reproducing in detail
the main system parameters of each of the 44 LFI channels \textquotedblleft as built \textquotedblright,
including total gain, noise temperature, bandpasses, non-linearity of the
response. The Real-Data LARFM has been successfully used to reproduce the LFI
behaviour observed during the LFI ground-test campaign \cite{2009_LFI_cal_M3,2009_LFI_cal_M4}.
Currently, the Real-Data LARFM is an essential element in the database of LFI
data processing centre and is available for detailed studies of
radiometer behaviour during the Planck survey.

%
%________________________________________________________________

%________________________________________________________________

\section{The {\sc Planck} LFI design}

%
%________________________________________________________________

\subsection{The overall LFI configuration}
\begin{figure}
	\centering
	\includegraphics[width=0.6\textwidth]{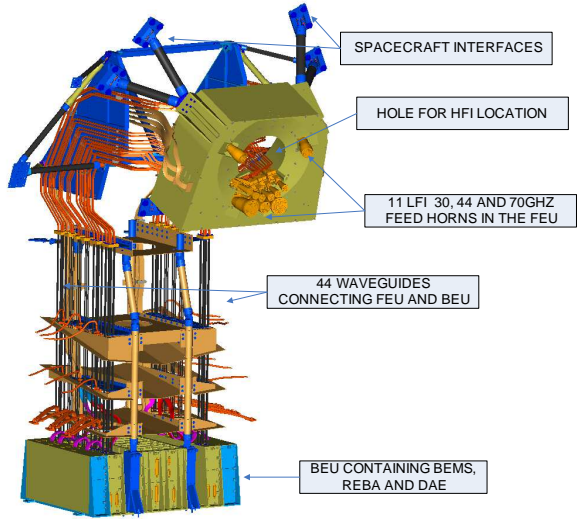}
	\caption{A drawing of the LFI with its spacecraft interfaces.}
	\label{figure1}
\end{figure}

LFI (\figurename~\ref{figure1}) is an array of 11 horns, each feeding 2 orthogonal linearly polarised channels
centred at 30, 44 and 70 GHz, on the focal plane of an aplanatic gregorian telescope. 
The LFI radiometers include the Front End Unit (FEU) and the Back End Unit (BEU), connected via 44 rectangular waveguides.
Such a division into FEU and BEU has been performed in order to minimise the effects of power dissipation in the critical focal
plane area. 

The Front End is cooled down to $20 K$ by one of two Sorption Coolers, while the Back End, which is
located in the body of the satellite, is at ambient temperature. The Front End Unit is the heart of the LFI instrument
and it contains the feed array and associated OrthoMode Transducers (OMTs), hybrid couplers and amplifiers blocks. The
Back End Unit comprises the Back End Modules (BEM) and the Data Acquisition Electronics.

%------------------------------
%------------------------------
%
%________________________________________________________________

\subsection{The LFI radiometer design}
\begin{figure}
	\centering
	\includegraphics[width=0.40\textwidth]{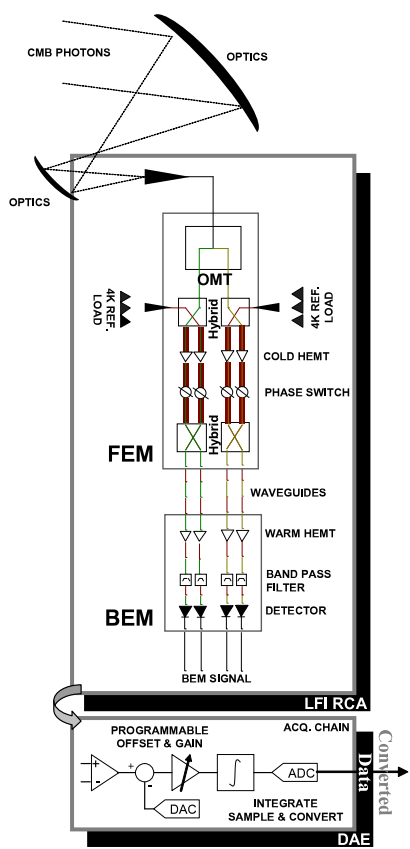}
	\caption{The Low Frequency Instrument Radiometric Chain Assembly (LFI RCA) with the Data Acquisition Electronics diagram.}
	\label{figure2}
\end{figure}

The LFI array is made up of eleven Radiometric Chain Assembly (RCA) (\figurename~\ref{figure2}) each incorporating the
two radiometers connected to the feed horn. Each radiometer includes a pair of amplification/detection chains correlated
through a pair of hybrid couplers, constituting a continuous-comparison device \cite{2009_LFI_cal_M2}. In this scheme,
the difference between the inputs to each of the chains (the signal from the telescope and that from a reference blackbody
load, respectively) is continuously being taken. This scheme strongly suppresses 1/f noise generated by instabilities in the RF
amplifiers, but not in the receiver elements following the second hybrid coupler. To remove this Back End 1/f, it is
necessary to modulate the sign of the comparison using solid-state phase-shifters within the correlation section. The
differencing receiver improves the stability at a cost of a factor of \begin{math}\sqrt{2}\end{math} in sensitivity compared
with a total power scheme. Any difference between the sky and reference input signals can give rise to 1/f noise, in the
LFI receivers this is further reduced by scaling one of the output data streams by a gain modulation factor "r". The
blackbody reference itself must remain at a very stable temperature (see \cite{2009_LFI_cal_R1}).

%------------------------------
%------------------------------

The intensity $\Delta$T of the smallest detectable signal is given by the radiometer equation (from \cite{Skou}):

%------------------------------
\begin{equation}
	\Delta T = \sqrt{2}\ \frac{T_{\mathrm{in}} + T_{\mathrm{noise}}}{\sqrt{B_{\mathrm{eff}} \cdot \tau}},
\end{equation}
%------------------------------

where $T_\mathrm{in}$ is the input temperature to the radiometer, $T_\mathrm{noise}$ is the radiometer noise temperature and
$\tau$ is the integration time. Radiometers are characterised by a spectral response g($\nu$) through the radiometer
band. The effective bandwidth is defined as:

%------------------------------
\begin{equation}
	B_{\mathrm{eff}} = \frac{\left[ {\int\limits_0^\infty {g\left( \nu \right)d\nu}} \right]^2}
	{\int\limits_0^\infty {g^2 \left( \nu \right)d\nu}}.
\end{equation}
%------------------------------

%________________________________________________________________

\section{The analytical LFI Advanced RF model}
\label{sec:analytical}

The need to analyse the effect of LFI non-ideal response and to estimate the impact of systematic effects
lead to the development of a radiometer system simulator \cite{thesis_battaglia}, using the Agilent Advanced Design System
(ADS) software. ADS provides Radio Frequency (RF) models for development of system specifications. It contains an RF
simulator that predicts performance of complete RF systems and includes a set of block-level RF models for linear and
non-linear components. The ADS Design Environment, Data Display, RF System Simulator and Models tools enable the
graphic implementation of the receiver components and the possibility to obtain information about the RCA main properties,
such as system gain and noise temperature.

%
%________________________________________________________________

\subsection{Implementation}

The radiometer model is built as a network whose electromagnetic properties can be expressed using scattering
parameters, taking into account the losses and reflections of the input signal, when it passes through each network element.

For a generic multi-port network definition, it is assumed that each of the ports is allocated an integer n ranging
from 1 to N, where N is the total number of ports. The insertion loss is the loss in load power due to the insertion
of a component at some point in a transmission system. In measured insertion loss tables found in literature, the
\begin{math}\left|{S_{21}}\right|^{2}\end{math} is given rather than the pure insertion loss value. In order to evaluate
the pure insertion loss, the measured return loss has to be taken into account. The
\begin{math}\left|{S_{21}}\right|^{2}\end{math} is measured experimentally, in decibel, as

%------------------------------
\begin{equation}
	\left| {S_{21} } \right|^2 = \frac{{P_{out} }}{{P_{in} }}.
\end{equation}
%------------------------------

$S_{21}$ includes both the effect of the insertion and the return loss, in fact:

%------------------------------
\begin{equation}
	S_{21} = 10^{\frac{{IL_{dB} }}{{20}}} \sqrt {1 - 10^{\frac{{RL_{dB} }}{{10}}} },
\end{equation}
%------------------------------

and the pure insertion loss is given by:

%------------------------------
\begin{equation}
	IL_{dB} = \;10\;Log_{10} \left( {\frac{{\left| {S_{21} } \right|^2 }}{{1 - 10^{\frac{{RL_{dB} }}{{10}}} }}} \right).
\end{equation}
%------------------------------

Return loss, instead, is given by:

%------------------------------
\begin{equation}
	RL_{dB} = \;10\;Log_{10} \;\left| {S_{11} } \right|^2 ,
\end{equation}
%------------------------------

and total system gain is obtained by:

%------------------------------
\begin{equation}
	G\left( {dB} \right)\ = 10\;Log_{10} \left( {\frac{{P_{out} }}{{K_B \;\Delta \nu \left({T_{noise} + T_{in}}\right)}}} \right),
\end{equation}
%------------------------------

where $T_\mathrm{in}$ is the input load temperature and $P_\mathrm{out}$
the corresponding power at detector output; $\Delta\nu$ is the effective bandwidth and $K_\mathrm{B}$ is the Boltzmann's
constant.
$T_\mathrm{noise}$, the system noise temperature is the input load temperature necessary to generate, in a noiseless system, the same output voltage of the noisy device with zero input load.

%The noise temperature of a two-ports device characterised by its physical temperature $T_\mathrm{phys}$, its losses (IL)
%and reflections (RL), is given by:
%
%%------------------------------
%\begin{equation}
%	T_n = \left( {1 - 10^{\frac{{RL}}{{10}}} } \right)\left( {1 - 10^{\frac{{IL}}{{10}}} } \right)T_{phys},
%\end{equation}
%%------------------------------
%
%and, assuming no signal at its input port ($T_\mathrm{in}$ = 0 K), the system noise temperature $T_\mathrm{sys}$ is
%defined as:

%------------------------------
%\begin{equation}
%	T_{sys} = \frac{T_n}{{\left|S_\mathrm{21}\right|}^{2}},
%\end{equation}
%------------------------------

%i.e., $T_\mathrm{sys}$ is the noise temperature that, at the input of an ideal noiseless device, generates the same output
%noise temperature $T_\mathrm{n}$ of the noisy device.

%
%________________________________________________________________

\subsubsection{Devices implementation}

\begin{figure}[ht]
	\centering
	\includegraphics[angle=90,height=\textheight]{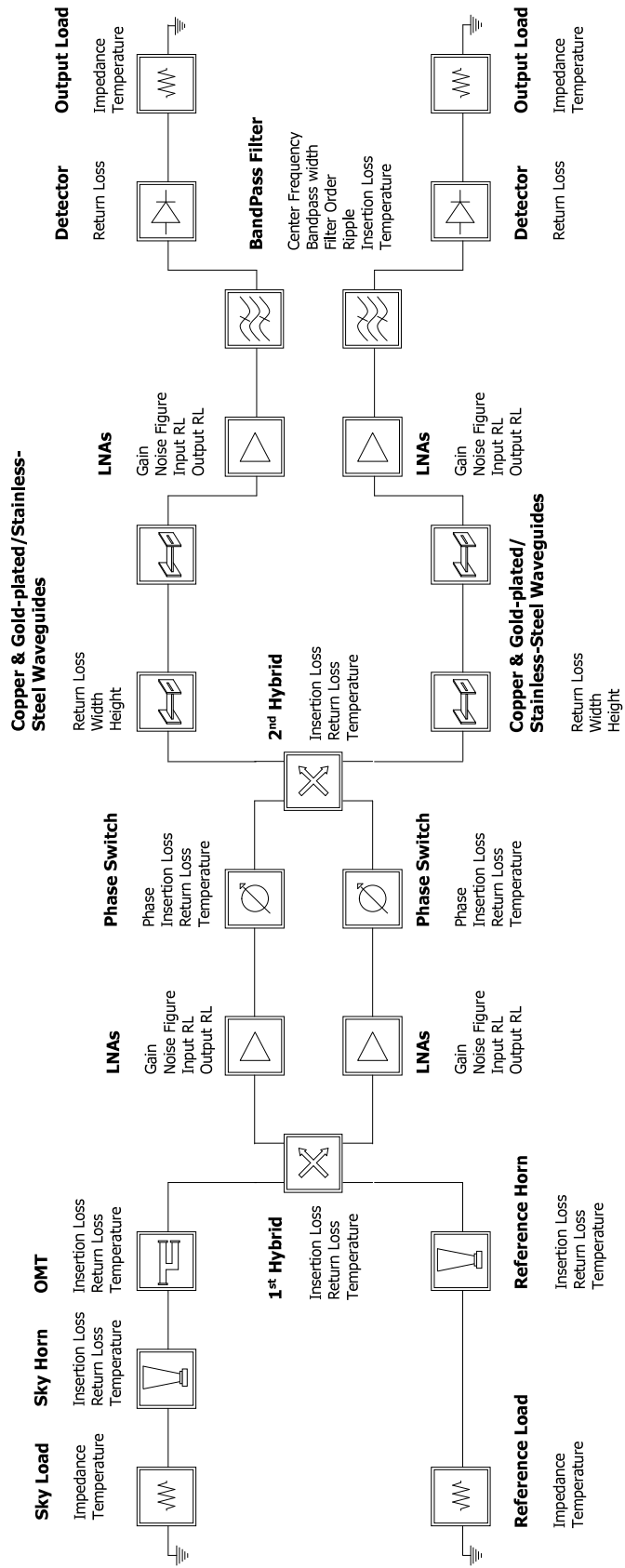}
	\caption{The Analytical LARFM graphical representation in the ADS design environment. Each component is a port device
	completely characterised by the parameters displayed in the figure.}
	\label{analytic_model}
\end{figure}

\figurename~\ref{analytic_model} shows a schematic of the implementation of the Analytic LARFM. The elements that constitute
the RCA receiver shown in \figurename~\ref{figure2} have been modelled in ADS starting from n-port device models, amplifiers, hybrids, wave-guides models and equation-based models.

%------------------------------
%------------------------------
\clearpage

\textbf{Input Loads}. The receiver input load is generated by the port component \textit{termination}. It is a $50 \Omega$
resistance at temperature $T_{ant}$ that produces a power noise equal to $P = K_BT_{ant}\Delta\nu$, where, in order to simulate the power absorbed by the sky and reference feed horns,
$T_{ant}$ is the antenna temperature related to the thermodynamic temperature $T$ by:

\begin{equation}
	T_{ant} = \frac{\frac{h\nu}{kT}}{e^\frac{h\nu}{kT} - 1}T.
\end{equation}

\textbf{Feed Horn and OMT}. The sky and reference feed horns, as well as the OMT, are modelled as two-port devices,
completely characterised by their scattering parameters - expressed in terms of insertion loss and return loss - and by their
physical temperature, taking into account the noise temperature $T_{noise}$ generated by the component itself into the network.
Given an input signal $T_{in}$ at the input port of, e.g., a feed horn, the signal at the output of the element is given by:

\begin{equation}
	T_{out} = \left|S_{21}\right|^2 T_{in} + T_{noise}. 
\end{equation}

\textbf{Hybrids}. The hybrid is modelled be a four port S-parameter set of equations using two values of magnitude and phase. They are set
to correctly reproduce the in-phase and anti-phase addition of the inputs. The hybrid model by default has a perfect port isolation, but
we tested also the effect of a crosstalk between the two paths of the signal (that simulates a non-ideal isolation between the two channels of the radiometer).
The hybrid, being a noisy component, is also characterised by the physical temperature of the device.

\textbf{Front End and Back End Low Noise Amplifiers}. The Low Noise Amplifier (LNA) is modelled using the amplifier component
of the ADS object library. The LNA model uses scattering parameters for modelling the amplifier gain ($S_{21}$) and reflections 
at the input and output ports of the device ($S_{11}$ and $S_{22}$); minimum noise figure ($NF_{min}$) at optimum source reflection
($S_{opt}$) and equivalent noise resistance $R_n$ are used instead of noise figure to characterise the amplifier noise, providing
also a correct return loss modelling response\footnote{http://edocs.soco.agilent.com/display/ads2009/Amplifier2+(RF+System+Amplifier)}.

\textbf{Phase Switch}. The phase switch model is a two ports system. Scattering parameters are defined in polar coordinates, so that the change in phase
 is obtained by setting the $S_{21}$ parameter phase to 180 degrees. The
temperature parameter characterises the noise properties of the phase switch model, as for the feed horn and OMT models.

\textbf{Waveguides}. The RCA waveguides are modelled into two parts, the copper waveguide model and the gold-plated/stainless-steel
waveguide model. Given the high thermal conductivity of copper, the copper waveguide model consists in a single rectangular waveguide
element at the same temperature of the Front End Module, i.e. 22K. At the input and output ports two impedance transformers play a
double role: they enable the waveguide matching to the characteristic impedance of the network (50 $\Omega$) and they model the waveguide
flanges return loss.\\
The gold-plated and stainless-steel waveguide model, although characterised by the same parameters, is more complex than the copper one,
including several rectangular waveguide elements, taking into account the linear temperature gradient between the two flanges.

\begin{figure}
	\centering
	\includegraphics[width=0.80\textwidth]{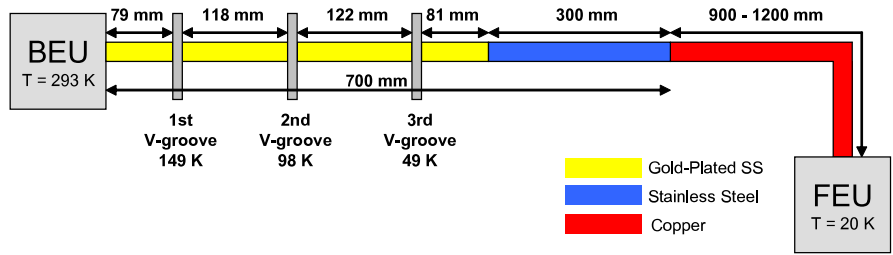}
	\caption{Schematic of the temperature distribution along the waveguides. V-grooves are thermal shields that isolate the cold Front
	End (22K) from the warm Back End Unit (293K).}
	\label{wg_schematic}
\end{figure}

In fact the stainless steel section interfaces to two heat sinks at either end, one at 300 K and the other at 22 K. The thermal
distribution along the stainless steel waveguide is critical due to its high electrical resistivity. So, with reference to
\figurename~\ref{wg_schematic}, the stainless steel section has been modelled so that the temperature at the flanges and shields
(V-groove) interfaces is preserved. For the gold plated sections the value for gold resistivity has been used.

By setting the dimensions (width and height) of the waveguides section, the temperatures at the interfaces and the return loss at the
flanges, the model is able to reproduce a cryogenic response of the waveguides, in terms of the insertion loss in the simulated frequency
band. 

\textbf{Bandpass filter}. Its design is a Chebyshev response bandpass filter and its parameters (see \figurename~\ref{analytic_model}
for a complete list) completely constrain the insertion loss in the passband and the slope at the cut-off frequency. From the noise
temperature analysis point of view, the band pass filter behaves like an attenuator, such as OMT and feed horns.

\textbf{Detector}. The detector has been modelled with a two ports system characterised by a return loss value. The power conversion
and integration over the band are performed by an appropriate calculation envelope, available in ADS, so that the output voltage
integrated over the frequency band $\Delta\nu$ of the radiometer is given by:

\begin{equation}
	V_{out} = \int_{\Delta\nu} G \cdot P_{out}(\nu)~d\nu,
\end{equation}

where G is a gain factor and $P_{out}(\nu) = V_{out}(\nu)^2/R$, being $V_{out}(\nu)$ the simulated output voltage as a function of the
frequency and $R$ = 50 $\Omega$ the impedance of the network.

Running ADS simulations it is possible to calculate:

%------------------------------
\begin{itemize}
	\item scattering parameters over the band for the whole RCA model and component by component;
	\item output voltage over the band;
	\item integrated output voltage.
\end{itemize}
%------------------------------

%
%________________________________________________________________

\subsection{Model verification}
The model is verified first at single component level and then at RCA level.
%
%________________________________________________________________

\subsubsection{Component level verification}
The functional consistency of each radiometer component is verified evaluating scattering parameters, output voltage
and noise temperature. At single component level the S-parameters are verified by the ADS primitive models themselves,
so that a verification of sub-groups of cascaded elements has been carried out, e.g. cascaded feed horn and OMT.
The simulated scattering parameters exactly match with those calculated using the scattering matrix theory, giving
confidence that ADS applies the same expressions to compute scattering parameters for a cascaded system.

\begin{figure}
	\centering
	\includegraphics[width=0.60\textwidth]{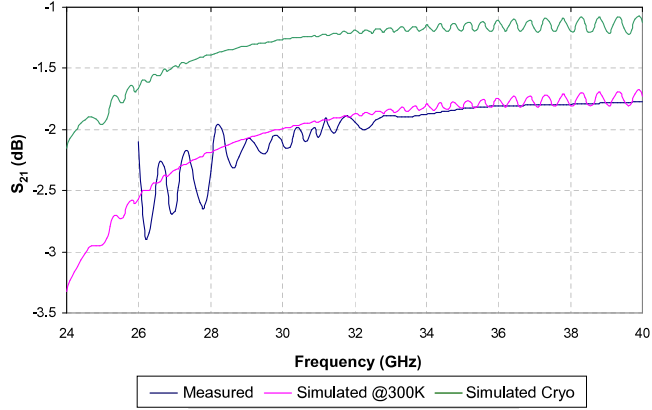}
	\caption{30 GHz waveguide $S_{21}$ at ambient temperature (measured and simulated) and cryogenic temperature
	(simulated only). Ripples in measurements are due to reflections within the test set-up; those in simulations
	are due to the impedance mismatch at the flanges interfaces, since -30 dB return loss is set in this simulation.}
	\label{wg_verif_all}
\end{figure}

Given the large quantity of cascaded elements, the validation of the waveguides model is performed by comparing the
simulation results with the measured $S_{21}$ at room temperature. Then a prediction of their behaviour at cryogenic
temperature is performed using the model, since no cryogenic measurements are available for the waveguides.
\figurename~\ref{wg_verif_all} show a comparison between simulation and measurements for the 30 GHz assembled waveguide
(copper section and gold-plated/stainless steel section) at ambient temperature (300K).

%------------------------------
%------------------------------

Cryogenic simulation shows that the $S_{21}$ is about 1dB better than that measured and simulated at ambient temperature.
The cryogenic model takes into account for the copper and gold-plated/stainless steel resistivity dependence on the
temperature distribution along the waveguides.\\

To evaluate the reliability of the model, functional tests and performance tests have been performed. Functional test
goals included calibration of the system and a check of the correct mechanism for phase switching.\\ The calibration
factor, $G$, giving the conversion of the diode's output signal (in volts) into physical units (in kelvins), is defined
as:

%------------------------------
\begin{equation}
	G = \frac{T_1 - T_2}{V_1 - V_2} \equiv \frac{\Delta T}{\Delta V},
	\label{eq_calibration}
\end{equation}
%------------------------------

where $\Delta$T is the variation between two different input temperatures and $\Delta$V is the corresponding variation
in the output voltages. From equation \ref{eq_calibration} it follows that the calibration factor $G$ is independent of
the temperature of the reference load.\\The following simulation set-up has been used to calculate the calibration factor:

\begin{itemize}
	\item Reference load temperature is fixed to 4K;
	\item Sky load temperature is changed between 20K and 77K;
\end{itemize}

\begin{figure}
	\centering
	\includegraphics[width=0.45\textwidth]{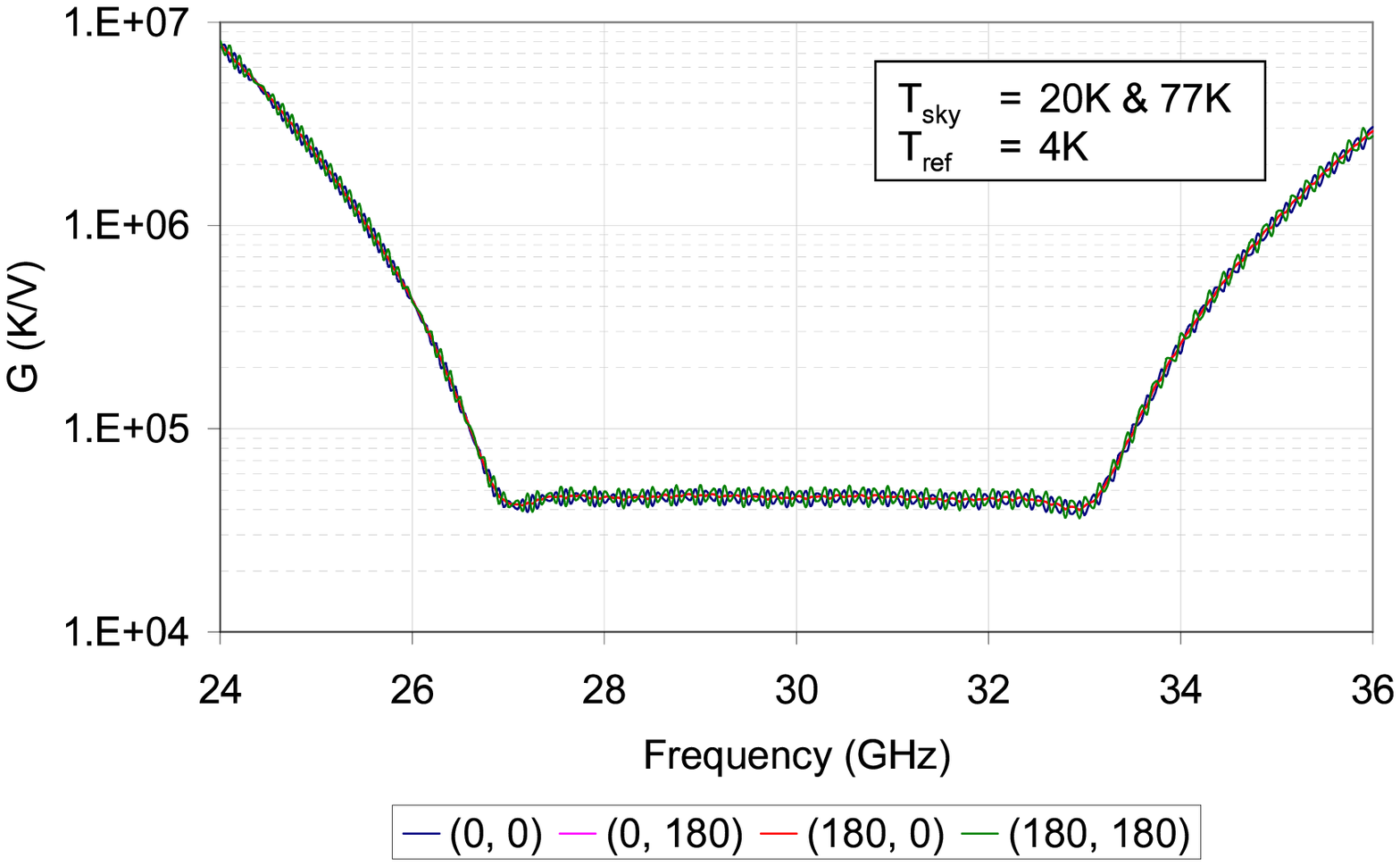}
	\includegraphics[width=0.45\textwidth]{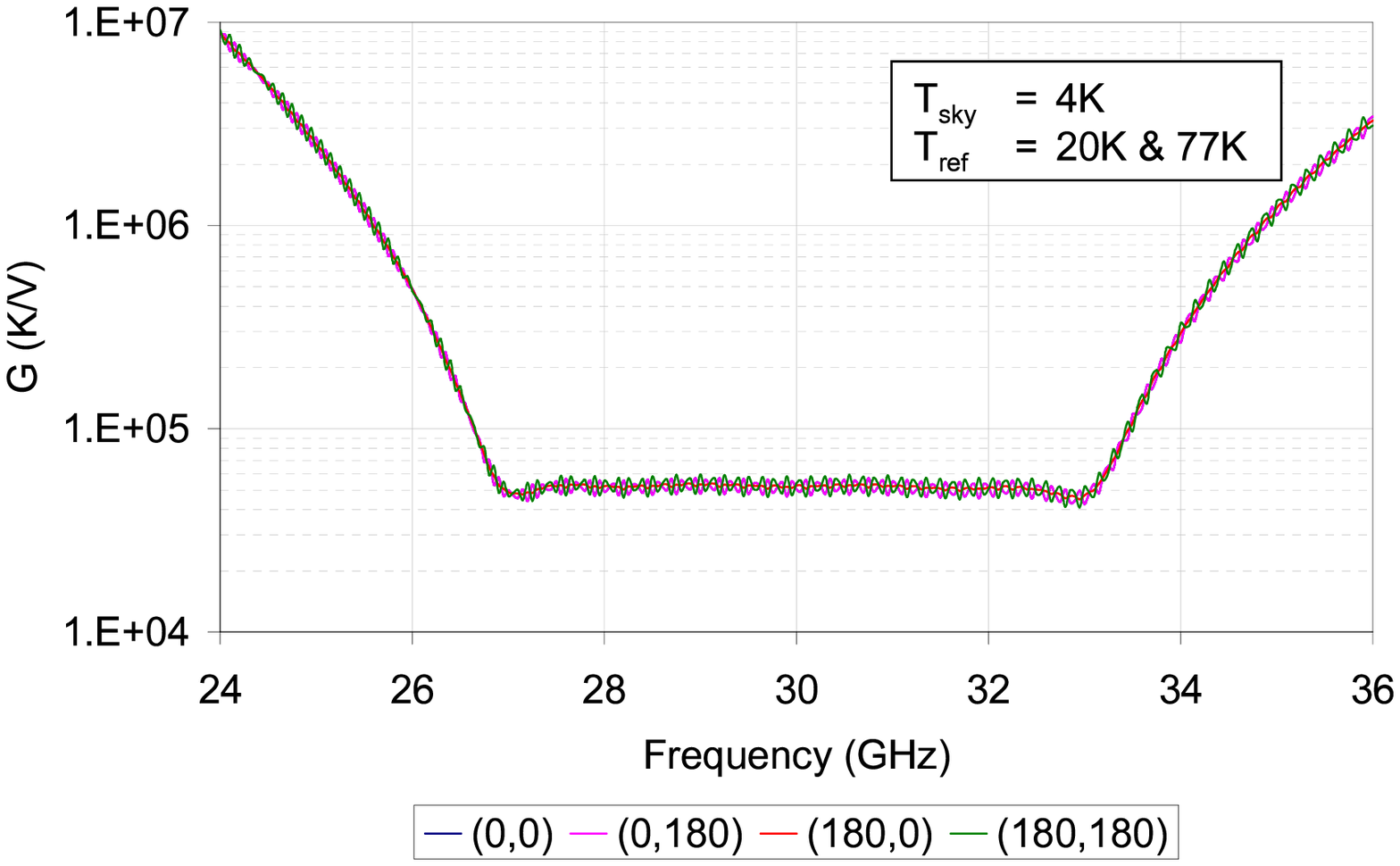}
	\caption{Calibration test results at 30 GHz. The factor $G$ is plotted in K/V versus frequency for fixed sky (left)
	and reference (right) load temperatures. The calibration factor is independent of phase switch status, labelled
	as ($Lag_1$, $Lag_2$).}
	\label{rca_calibration}
\end{figure}

All Phase Switches conditions have been analysed: both radiometer branches with 0 degree lag; both branches with 180
degree lag; the intermediate status in which only one branch signal is 180 degree shifted (labelled (0, 180) and (180
, 0) in \figurename~\ref{rca_calibration}). Note that the effect of the lag on a single branch of the radiometer is
to reverse the sky and reference signal at the outputs of the radiometer.
The simulation has been repeated by fixing the sky load temperature to 4K and varying the reference load between 20K
and 77K. \figurename~\ref{rca_calibration} shows the calibration factor $G$ in the 30 GHz RCA frequency band, in any
status of the two phase switches. Minor differences between the two results are due to asymmetries in the RCA (presence
of the OMT in the sky input).

%------------------------------
%------------------------------

%
%________________________________________________________________

\subsubsection{System level verification}
\label{sec:system_level}
\begin{figure}
	\centering
	\includegraphics[width=0.45\textwidth]{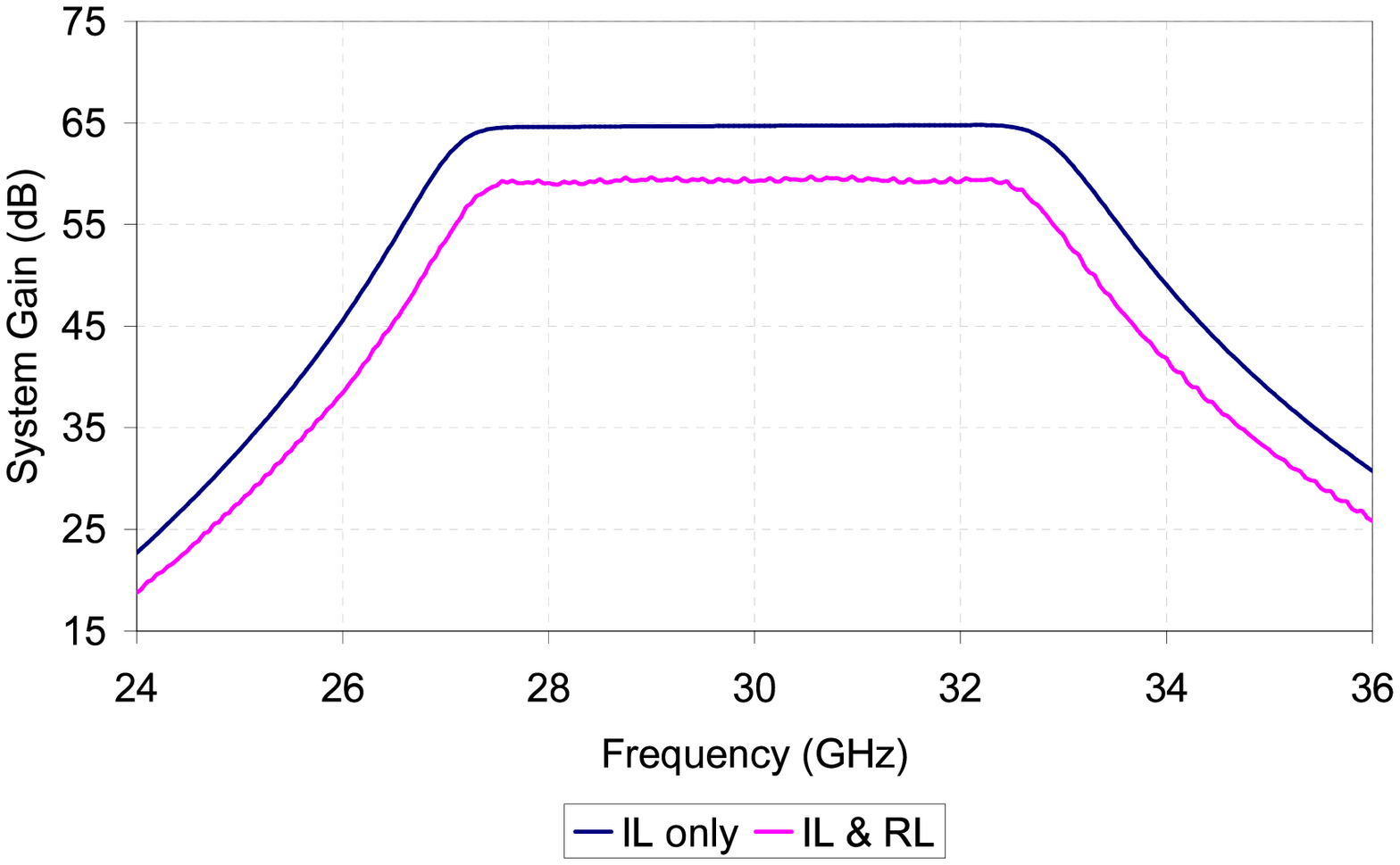}
	\includegraphics[width=0.45\textwidth]{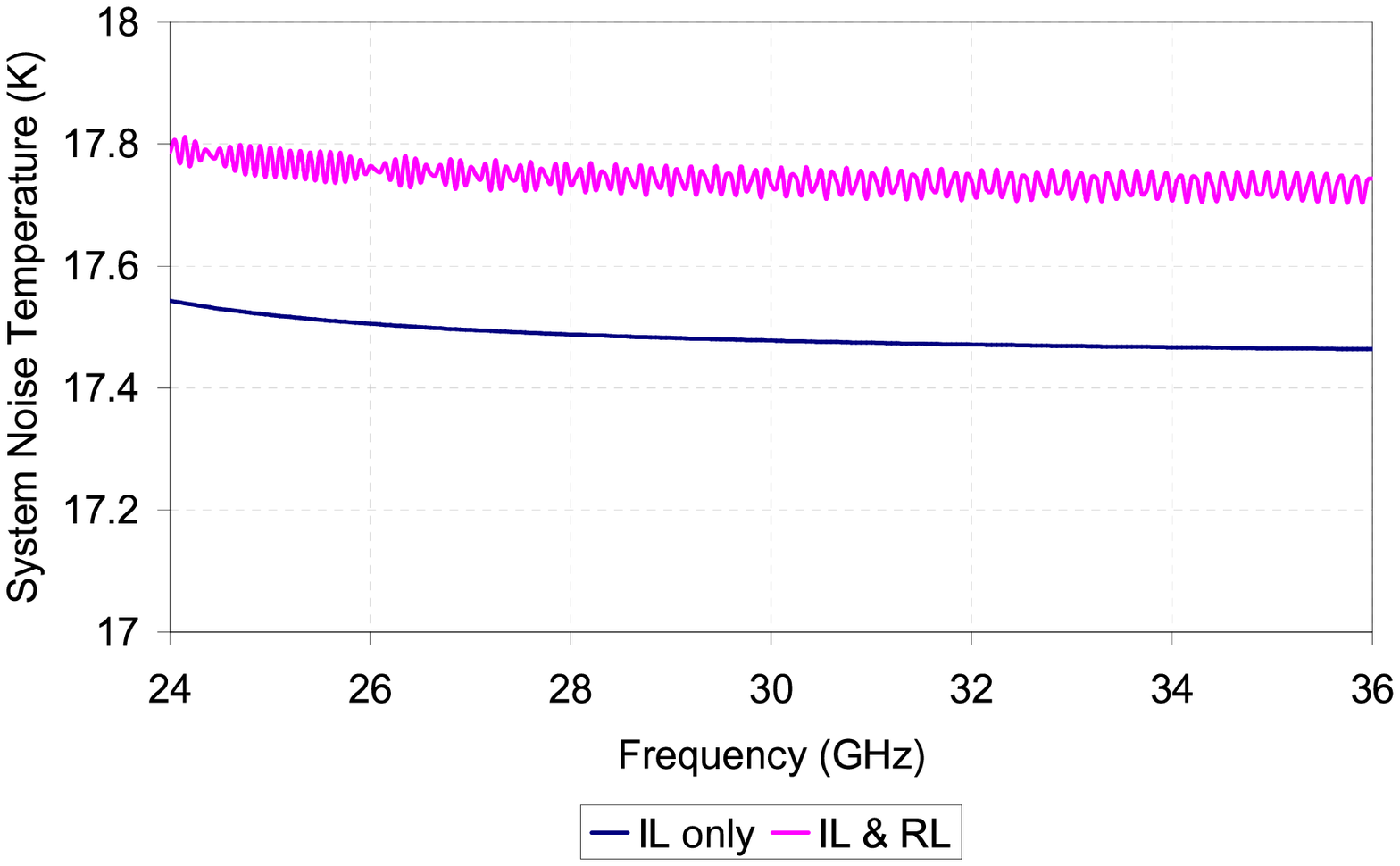}
	\caption{System Total Gain and System Noise Temperature of a 30 GHz RCA characterised by insertion loss only and
	both insertion loss and return loss. When reflections are calculated by the simulator, the total gain is reduced
	and the system noise temperature increases.}
	\label{performance}
\end{figure}

Performance tests have been conducted, including system gain and system noise temperature calculus in the LARFM. All
components parameters are set to their nominal value as from the {\sc Planck} LFI specifications and requirements.
The case of an RCA characterised by the insertion loss only is simulated; then simulations are repeated by taking into
account for the reflections of the input signal (return loss).
(\figurename~\ref{performance}) shows a generally good agreement with expected behaviour. Reflections cause a general
fall of the system gain value and the formation of band ripples, which contribute to an effective bandwidth reduction.
System noise temperature is also affected by the presence of reflections at the waveguides flanges, causing ripples;
moreover its mean value in the frequency band increases to about 1.5\%.

%------------------------------
%------------------------------

%
%________________________________________________________________

\subsection{Applications}

\subsubsection{Instrument performances with two FEM prototypes}
\begin{figure}
	\centering
	\includegraphics[width=0.45\textwidth]{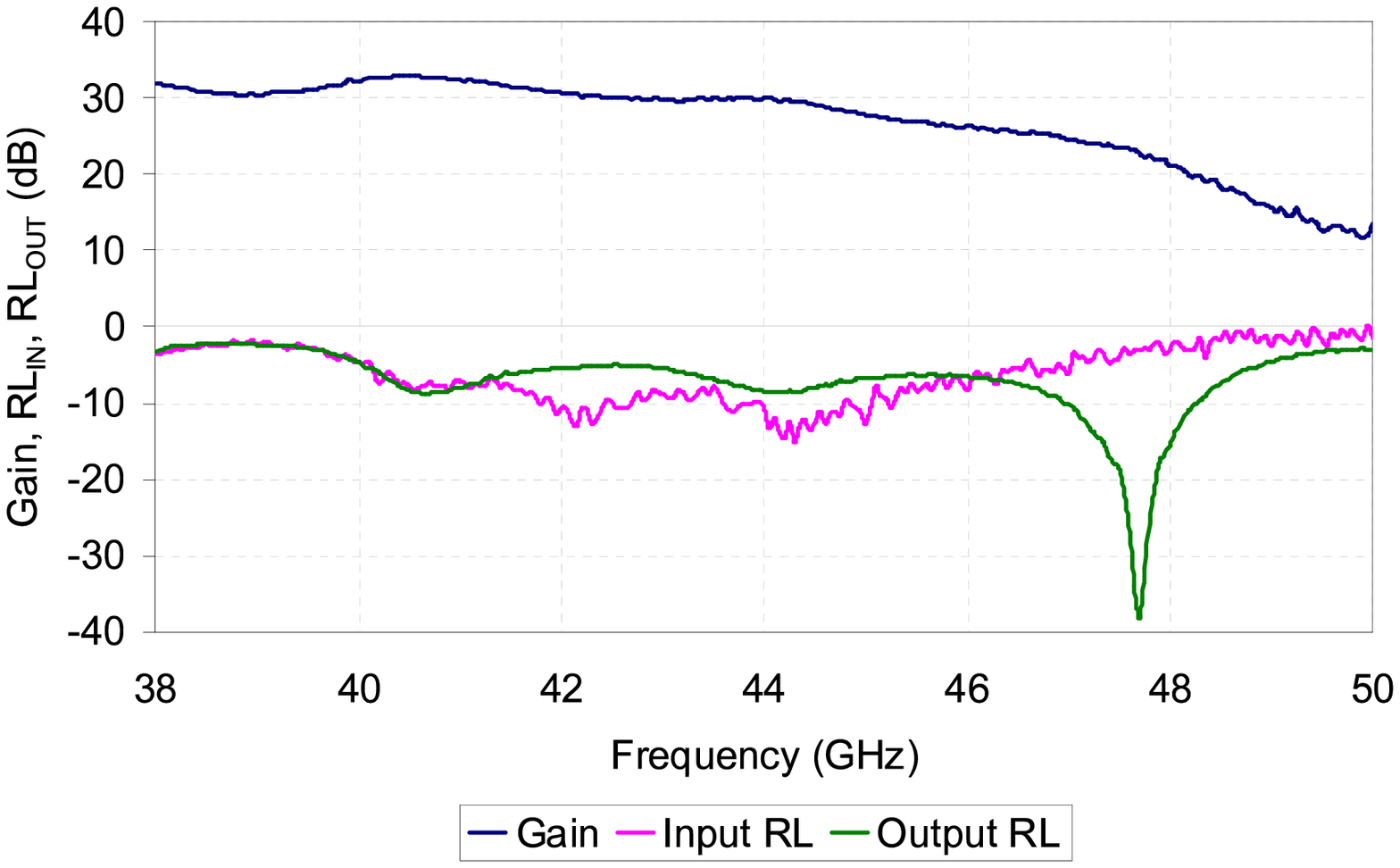}
	\includegraphics[width=0.45\textwidth]{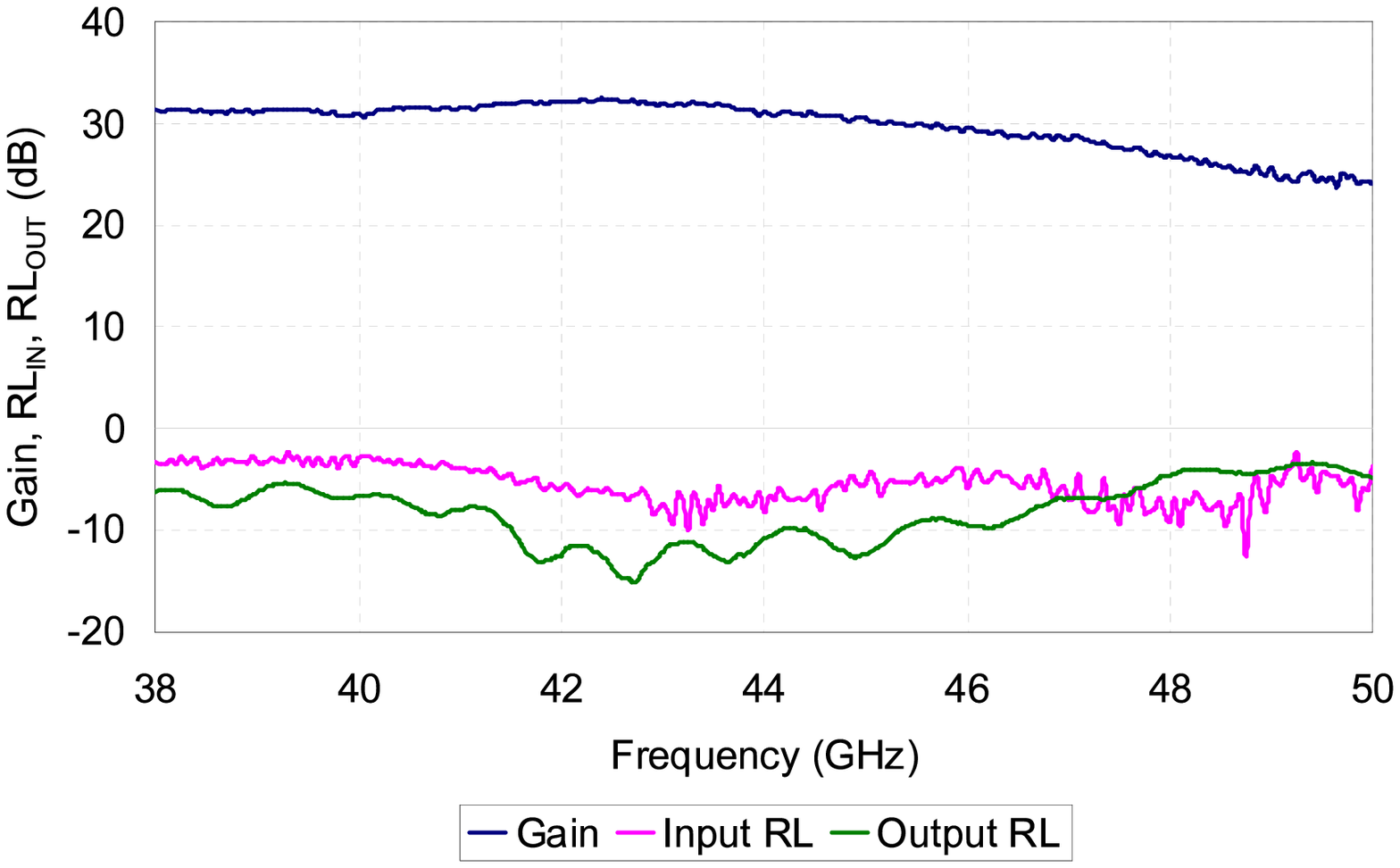}
	\caption{(Left) The first 44 GHz LNA prototype performances: input return loss, output return loss and gain measured
	at 293K. (Right) The second 44 GHz LNA prototype measurements at 295K.}
	\label{prototypes_data}
\end{figure}

\begin{figure}
	\centering
	\includegraphics[width=0.45\textwidth]{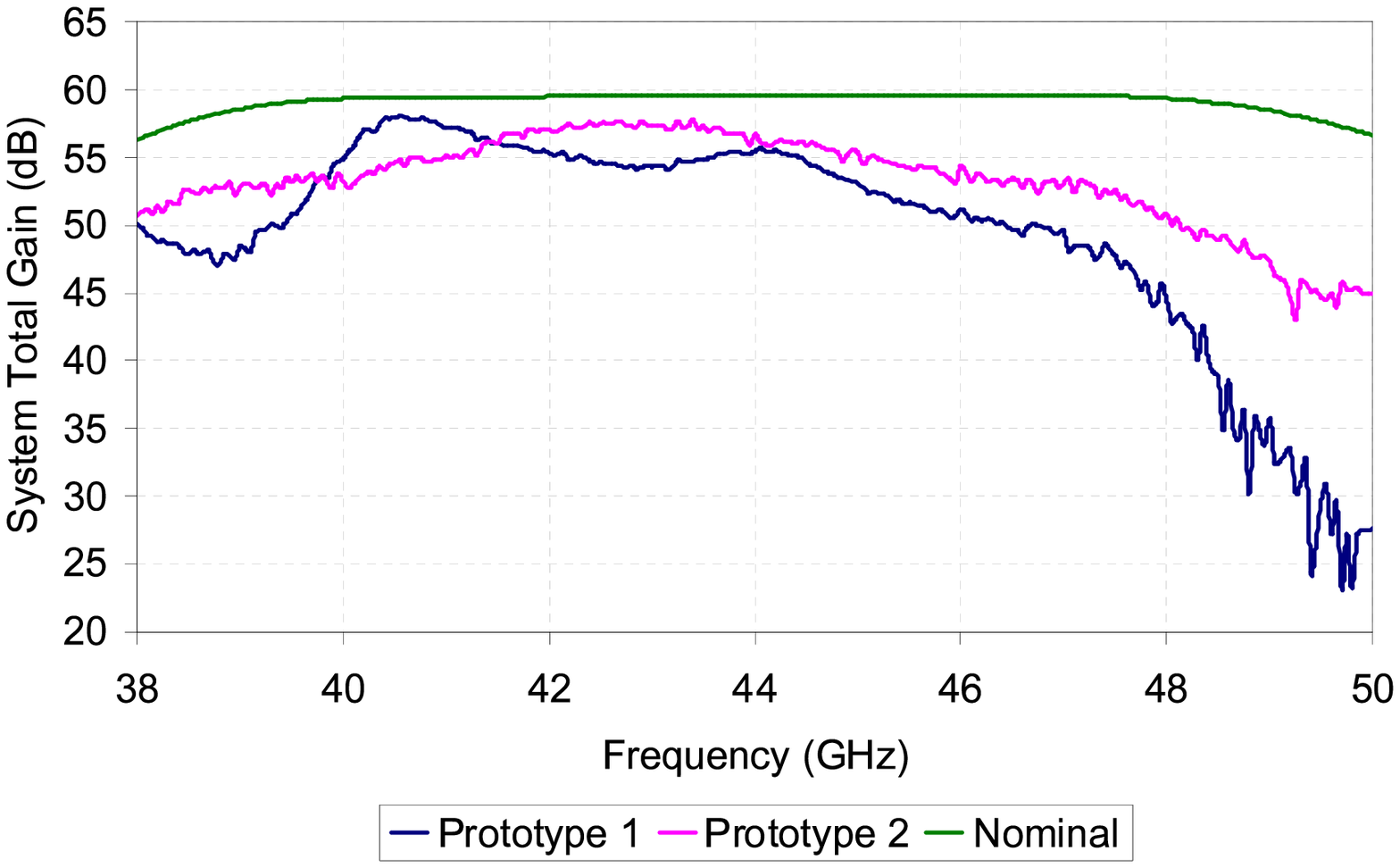}
	\includegraphics[width=0.45\textwidth]{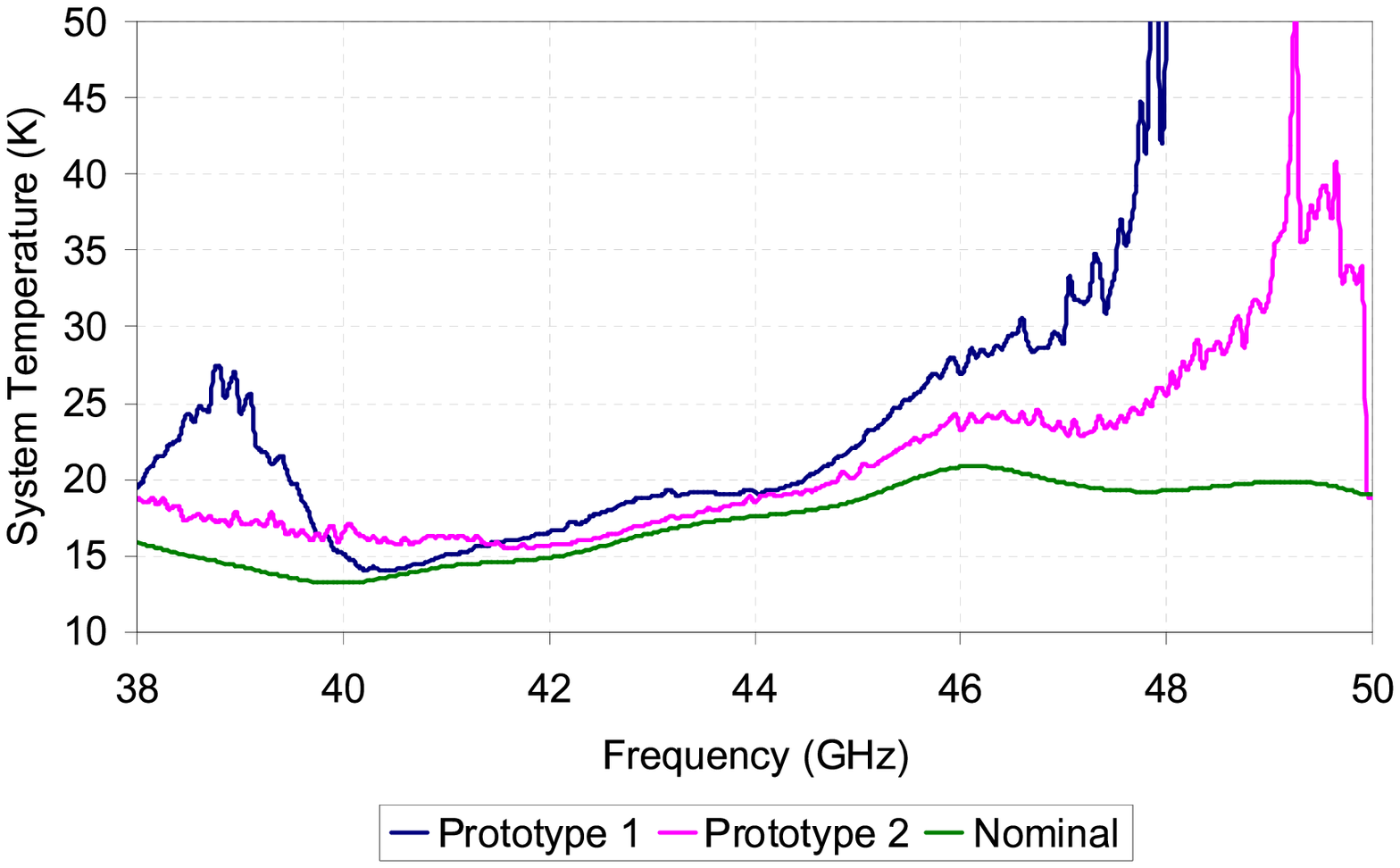}
	\caption{System Total Gain and Noise Temperature simulated for 44 GHz RCA's including two LNA prototypes and the
	nominal case (LNA performances as from requirements).}
	\label{prototypes_results}
\end{figure}

As a first application of the LARFM to the analysis of the instrument performances \cite{thesis_franceschet}, we evaluated
the system response in the frequency band in two different configurations including two prototypes of the Front End amplifiers
at 44 GHz, whose performances, in terms of input/output return loss and total gain are measured in the frequency range
38--50 GHz (\figurename~\ref{prototypes_data}). The LNA noise figure is given by the measurements on a third LNA and has not
changed during the entire analysis of the prototypes; its profile is visible in System Noise Temperature simulation for the
nominal case (see \figurename~\ref{prototypes_results}, right).

%------------------------------
%------------------------------

In order to evaluate the effect of the Front End LNA performance on the system total gain, noise temperature and
effective bandwidth, we set all RCA parameters, with the exception of those of the Front End amplifiers, to their
nominal value at 44 GHz. The in-band performances of the two LNA prototypes are measured at ambient temperature (a worst
case). The nominal case, with the LNA input return loss $RL_{in}$ = -5 dB, output return loss $RL_{out}$ = -6 dB and gain
$G$ = 30 dB, is also simulated for a comparison.

\figurename~\ref{prototypes_results} shows that in both simulations including LNA prototypes, the total gain follows
the shape of the corresponding prototype, with wide ripples for frequencies above 47 GHz, due to the effect of the input
return loss. As expected, the RCA with the second LNA prototype exhibits better performance in terms of total system gain,
in particular for frequencies above 44 GHz. Simulations show similar results for the system noise temperature: it rapidly
diverges for frequencies above 48 GHz in the first prototype simulation; the input return loss of the LNA plays a leading
role in this simulation, since nearly all incoming power is reflected at the LNA input port.

%------------------------------
%------------------------------

The effective bandwidth, as calculated by the LARFM for all test cases is summarized in \tablename~\ref{prototypes_bw}.

%------------------------------
\begin{table}[h]
	\centering
	\begin{tabular}{c c c}
		\hline\hline
		\textit{Frequency Range} & \textit{38--50 GHz} & \textit{39.6--48.4 GHz}\\
		\hline
		\textit{$B_{eff}$ Prototype 1} & 6.93 & 6.28\\
		\textit{$B_{eff}$ Prototype 2} & 8.94 & 7.45\\
		\textit{$B_{eff}$ Nominal} & 11.75 & 8.84\\
		\hline\hline
	\end{tabular}
	\caption{Effective bandwidth $B_{eff}$ in the measured data frequency range (38--50 GHz) and in the {\sc Planck} LFI
	required frequency bandwidth (20\% of the centre frequency).}
	\label{prototypes_bw}
\end{table}
%------------------------------

This analysis shows that the LARFM is a useful tool to investigate the behaviour of the {\sc Planck} LFI instrument when
in band non idealities characterise the system components.

%
%________________________________________________________________

\subsubsection{Effects of OMT asymmetry}
If the OMT arms have different performance, in terms of loss and reflection in the frequency band, the two
orthogonal components of the polar signal are affected differently, and will be confused with a real polarisation signal
when processing the polarisation information. As an application of the LARFM, the impact of asymmetries, in terms of
losses and reflections, in each one of the OMT arms on the in-band total system gain is studied. Simulations are performed
by setting all RCA parameters to their nominal values, with the exception of the OMT insertion loss and return loss,
measured on the Qualification Model (QM) of the 30 GHz OMT.

\begin{figure}
	\centering
	\includegraphics[width=0.60\textwidth]{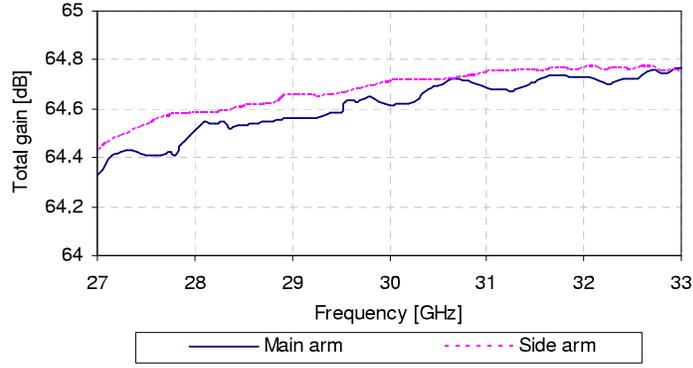}
	\caption{Effect of the OMT in band asymmetries on the total system gain of the 30 GHz RCA Qualification Model.}
	\label{omt_mismatch}
\end{figure}

Simulation results in \figurename~\ref{omt_mismatch} shows
that the maximum gain mismatch between the arms of the radiometer is about 0.25\% at 30 GHz.

%------------------------------
%------------------------------

\begin{figure}
	\centering
	\includegraphics[width=0.45\textwidth]{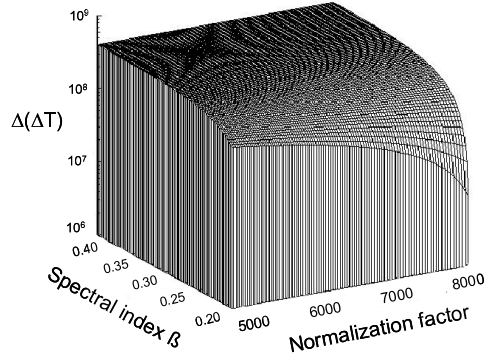}
	\caption{Difference, in mK, between the final output of the two branches of the radiometer.}
	\label{polar_dt}
\end{figure}

For a description of the impact of asymmetries between the two arms of the 30 GHz RCA on the CMB polarisation measurement,
the following expression has to be evaluated:

%------------------------------
\begin{equation}
	\begin{array}{l}
		\Delta \left( {\Delta T} \right) = \frac{1}{{\Delta \nu }}\left[ {\int {G_1 \left( \nu \right)T_{sky}
		\left( \nu \right)d\nu } - \int {G_1 \left( \nu \right)r_1 T_{ref} \left( \nu \right)d\nu } } \right] + \\ 
		\quad \quad \quad - \frac{1}{{\Delta \nu }}\left[ {\int {G_2 \left( \nu \right)T_{sky} \left( \nu 
		\right)d\nu } - \int {G_2 \left( \nu \right)r_2 T_{ref} \left( \nu \right)d\nu } } \right] \\ 
	\end{array}
\end{equation}
%------------------------------

where \begin{math}G_{1}(\nu)\end{math} and \begin{math}G_{2}(\nu)\end{math} are the gain relevant to the two arms of the
radiometer (\figurename~\ref{omt_mismatch}), $T_{sky}$ and $T_{ref}$ are the observed sky temperature and the reference load
temperature and $r_{1}$ and $r_{2}$ are the gain modulation factors for both arms of the RCA. The sky signal is
parametrised assuming a power-low behaviour for the sky temperature,

%------------------------------
\begin{equation}
	T_{sky} = \alpha \nu ^{ - \beta } 
\end{equation}
%------------------------------

where $\alpha$ and $\beta$ are derived from WMAP foreground maps at 23, 33, 41, 61 and 94 GHz and the CMB signal expression,

%------------------------------
\begin{equation}
	T_{CMB} \left( \nu \right) = \frac{{{{h\nu } \mathord{\left/
	{\vphantom {{h\nu } {K_B T_0 }}} \right.
	\kern-\nulldelimiterspace} {K_B T_0 }}}}{{e^{{{h\nu } \mathord{\left/
	{\vphantom {{h\nu } {K_B T_0 }}} \right.
	\kern-\nulldelimiterspace} {K_B T_0 }}} - 1}}T_0 
\end{equation}
%------------------------------

where $T_{0}$ = 2.725 K, and $K_{B}$ is the Boltzmann's constant. The K, Ka and Q band data fix the $\alpha$ and $\beta$
values in the ranges $\alpha \in [5\cdot10^{3}, 6\cdot10^{3}]$ and $\beta \in [0.25, 0.35]$. The reference load
temperature follows the frequency dependant 4.8K black-body curve for this analysis.\\By calculating the integrals for
the specified variation ranges of the normalisation parameter $\alpha$ and of the spectral index $\beta$, the difference
between the two differenced output signals, i.e. the expression $\Delta(\Delta T)$, is plotted in \figurename~\ref{polar_dt}.

%------------------------------
%------------------------------

%________________________________________________________________

\section{The Real-Data LFI Advanced RF Model}
\label{sec:realdata}

During the qualification and performance tests, each component of the LFI radiometers was independently characterised
in terms of frequency response. These hardware measurements replaced the parameters used in the analytical version of
the LARFM with the objective of estimating the effective response of the 44 channels of the LFI flight hardware.

The new LARFM maintains the same implementation just for the waveguides simulator, while the other units are replaced by their measurements.
Moreover, the main objective of the LARFM became the estimation of the bandpass response, see \cite{2009_LFI_cal_R3}, due to the high impact of systematic effects on end-to-end bandpass measurements.
The next sections explain the implementation and the results obtained with the LARFM.

\subsection{Implementation}

\begin{figure}[ht]
    \centering
    \includegraphics[width=\textwidth]{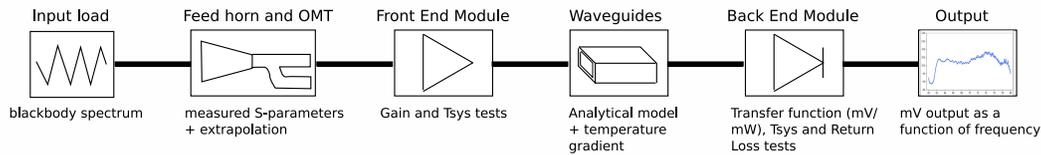}
    \caption{Model schematic for each LFI channel:
    feed horn and OMT are characterized by their S-parameters; FEM by gain and noise temperature; waveguides are simulated analytically; BEM by return loss, transfer function and integrated noise temperature.}
    \label{fig:model_schematic}
\end{figure}

\figurename~\ref{fig:model_schematic} shows a schematic of the implementation of the Real-Data LARFM. 
In this model the pseudo-correlation strategy is not considered, see section~\ref{sec:implemfem}, so each channel is completely independent from the other channel of the same radiometer.
Moreover, the number of devices is strongly reduced because FEM and BEM devices where fully characterized after assembly.

\subsubsection{Feed horn and OMT}

Feed horn and OMT return losses were measured on their assembly, injecting the test signal from the OMT port to be connected to the FEM. Insertion losses, instead, are available for the OMT only, but the impact of the feed horn is negligible.
Feed horn and OMT model is a single S\-parameters passive component based on these data.

\subsubsection{FEM}
\label{sec:implemfem}

A FEM model considering the pseudo correlation strategy needs spectral response measurements of each component
assembling the FEM:

\begin{itemize}
	\item Hybrids
	\item Low Noise Amplifiers
	\item Phase switches
	\item Internal FEM waveguides
\end{itemize}

However, all these components when tightly assembled into the FEM change their response significantly due to mutual interactions.
It is therefore more reliable to use the data from the tests of the complete FEM.

Each channel is simulated independently from the others using data from gain and noise temperature tests performed by the FEM producers.
Each channel is considered as a total power radiometer frozen in the configuration of the phase switches states which connects it to the sky signal. Indeed, there are two phase switch state combinations where the same BEM
channel is connected to the sky arm, but, after phase switch balancing, the responses match better than $\sim.1 dB$.

In ADS, the FEM has been modelled as an amplifier with gain given by the performance tests on each FEM
channel and noise figure computed by the system temperature as:

%------------------------------
\begin{equation}
	F = 10\log _{10} \left( {\frac{{T_{sys} }}{{290K + 1}}} \right).
\end{equation}
%------------------------------

FEM $S_{11}$ could only be measured at ambient temperature, therefore the LARFM implements the mean expected value $S_{11} = -7 dB$.
The RCA is then studied in an ideal configuration, in which all four RCA channels are looking simultaneously to the sky signal.

\subsubsection{BEM}
\label{sec:bemimplem}
The BEM model is made up of:

%------------------------------
\begin{itemize}
	\item Reflection only S\-parameter component for return loss
	\item A noise generator for the noise temperature
	\item An equation object for the gain
\end{itemize}
%------------------------------

In order to model the whole RCA, the RF signal coming from the FEM through the waveguides is summed with a noise
generator based on the measured BEM noise temperature. This signal is then amplified by the measured spectral transfer
function converting input power to output volts at a given frequency, which includes the effect of all the BEM components
together.

The output is a bandpass response of the complete RCA which can then be integrated over frequency in order
to compute the detector voltage output that can be compared with test results. 

During the flight hardware test campaign, 30 and 44 GHz RCAs showed signal compression at BEM level, i.e. the BEM output doesn't increase linearly with the input temperature. 
This effect has a big impact on system temperature calculation, because the data extrapolation process emphasises even a small non-linearity. Therefore, a parametric non linear analytic model was developed in \cite{2009_LFI_cal_R4} in order to fit test data for extracting the overall RCA gain, system temperature and BEM compression factor. The
model is based on the following equation:

%------------------------------
\begin{equation}
	V_{out} = G_0 \cdot \left[ {\frac{1}{{1 + b \cdot G_0 \cdot \left( {T_A + T_{noise} } \right)}}} \right] \cdot
	\left( {T_A + T_{noise} } \right),
\end{equation}
%------------------------------

where $T_{A}$ is the antenna temperature, $G_{0}$ is the linear gain and $b$ is the
compression factor. The compression factor is a single scalar value per channel which represents the BEM gain dependence
on the input power. BEM tests were performed at a power level where compression is expected. The first operation to
perform then is to calculate the linear gain by using the known information on the compression factor measure in the
relevant tests. The equation used is:

%------------------------------
\begin{equation}
	G_{lin} = \frac{{G_{test} }}{{1 - b \cdot P_{in} \cdot G_{test} }},
\end{equation}
%------------------------------

where $G_{test}$ is the total BEM gain or transfer function (i.e. output voltage vs input power), $b$ is the scalar
compression factor, $P_{in}$ is the input power for the BEM test and $G_{lin}$ is the computed linear gain. The last
step is to substitute $G_{lin}$ for the gain parameter of the BEM and change the equation describing the BEM to:

%------------------------------
\begin{equation}
	V_{out} = \frac{{G \cdot P_{in} }}{{\left( {1 + b \cdot P_{in} \cdot G} \right)}}.
	\label{eq:bemgain}
\end{equation}
%------------------------------

%
%________________________________________________________________

\subsection{Results}

The Real-Data LARFM outputs are the channel bandpasses, volt outputs, gains and noise temperatures either on the full RCA or in any section of it.
These data are the basis of several analyses on the LFI radiometers, in the following paragraphs we are going to introduce the most interesting.

\begin{figure}[h]
    \centering
    \includegraphics[width=.6\textwidth]{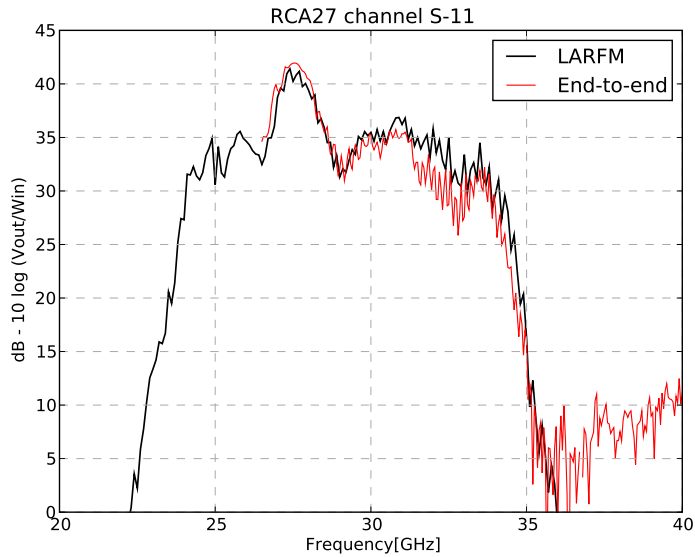}
    \caption{LFI27S-11 at 30 GHz. Comparison of LARFM bandpasses with end-to-end swept source measurements performed on WR28 nominal bandpass, between 26.5 and 40 GHz, see \cite{2009_LFI_cal_R3}.}
    \label{fig:lfi2711}
\end{figure}

The most important result of the LARFM is bandpass estimation, see \cite{2009_LFI_cal_R3}. \figurename~\ref{fig:lfi2711} shows  a good agreement between LARFM bandpasses and end-to-end measurements, the end-to-end measurements are however affected by systematic effects and their quality is not sufficient for a quantitative analysis.

\begin{figure}
	\centering
	\includegraphics[width=.6\textwidth]{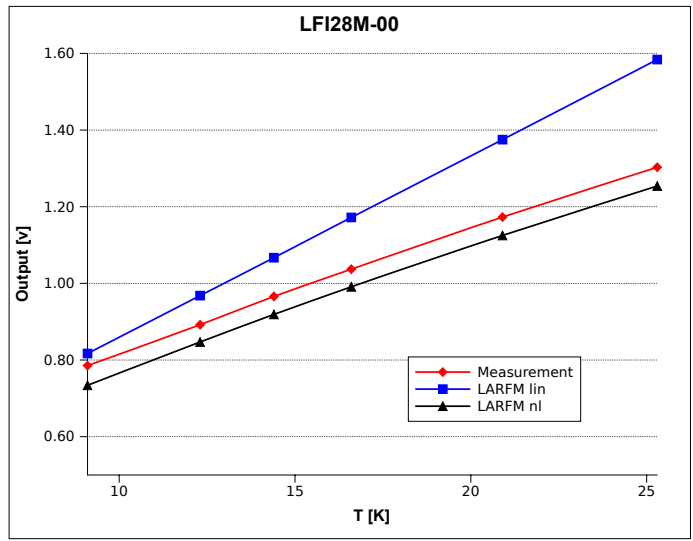}
	\caption{Output voltage versus temperature for a linearity test compared to the LARFM linear and non-linear BEM models.}
	\label{figure14}
\end{figure}

\figurename~\ref{figure14} shows the output voltage as a function of input load temperature measured in a linearity test compared to the LARFM results assuming a linear BEM response and implementing the BEM non-linear model as explained in section~\ref{sec:bemimplem}.
The compression parameter $b$ of Eq.~\ref{eq:bemgain} was measured using the test results and the comparison shows a good agreement between the non-linear model and the measurements.

%------------------------------
%------------------------------

\begin{figure}[h]
    \centering
    \includegraphics[width=.6\textwidth]{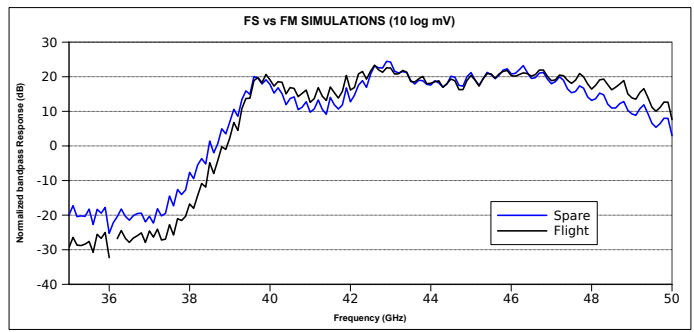}
    \caption{Comparison between LARFM bandpasses with nominal versus spare FEM on LFI24M-00}
    \label{fig:fsvsfm}
\end{figure}

The LARFM was also successfully applied to predict the impact on bandpasses of the substitution of a damaged FEM unit with a spare unit, to assess whether it would be acceptable to replace or necessary to repair the unit.
Figure~\ref{fig:fsvsfm} shows the impact on the LARFM bandpasses of the eventual substitution of the flight FEM with the spare unit on RCA 24. The response of the spare FEM is more structured than the nominal unit and the bandwidth is reduced by about $20\%$ (see \cite{phd_zonca}), the unit was successfully repaired.

The RF model has been used also during the cryogenic test campaign to successfully investigate the saturation at the BEM diode, by modelling the RF power level at the diode input. 

A more direct use of the model is foreseen in the unfortunate case of the radiometers working in non-nominal configurations, for example different thermal conditions.
In this case it is possible to estimate the impact of these non-nominal conditions on each component and then make use 
of the LARFM to simulate the non-nominal behaviour in terms of bandpasses and output voltage.
%________________________________________________________________

\section{Conclusions}
We have described the LFI Advanced RF Model, based on commercial software tools and measurements of the radiometer component
at single unit level. The LARFM provides a good description of the response of each of the 44 channels of the {\sc Planck}-LFI
array, such as radiometer response (including non-linear behaviour), system gain, noise temperature, spectral shape of the
radiometers bands. 

The measurements of the scattering parameters of the single units (i.e., feed horns, OrthoMode transducers, front-end
modules, waveguides, back-end modules) are very precise (typically better than 0.1dB), while in some cases the end-to-end
measurements of the integrated radiometer properties are more difficult. This is the case, for example, for the evaluation
of the shape of the spectral bands, whose knowledge is particularly important for polarisation analysis. In these cases,
the availability of the LARFM has proved particularly useful.

After being successfully used to support the {\sc Planck}-LFI design and ground-test campaigns, the model is now incorporated
in the LFI data processing centre to allow detailed studies of radiometer behaviour during the {\sc Planck} survey. Similarly,
future precision radiometric instruments will likely require similar models to be developed to support instrument development
and data analysis. Our experience has shown the usefulness of such an RF software model and indicates a strategy and some
technical tools that may be useful for other projects in the future.

%________________________________________________________________

\acknowledgments
Planck is a project of the European Space Agency with instruments funded by ESA member states, and with
special contributions from Denmark and NASA (USA). The Planck-LFI project is developed by an International Consortium
lead by Italy and involving Canada, Finland, Germany, Norway, Spain, Switzerland, UK, USA. The Italian contribution to
Planck is supported by the Italian Space Agency (ASI). The US Planck Project is supported by the NASA Science Mission
Directorate. In Finland, the Planck project is supported by the Finnish Funding Agency for Technology and Innovation
(Tekes). In the UK the Planck project is supported by PPARC (now STFC).

P. Battaglia, C. Franceschet and A. Zonca would like to thank Thales Alenia Space Italia S.p.A. Milano for having
supported the LARFM development in the framework of the PLANCK LFI experiment. P. Battaglia, C. Franceschet and A. Zonca
would like to thank all people working on PLANCK LFI at Universit\`a degli Studi di Milano, Istituto di Fisica del Plasma
-- CNR (Milano) and INAF -- IASF (Bologna) for their collaboration in model components definition and for providing RF
measurements data.\\A special thank to P. Meinhold and R. Hoyland for their revision and suggestions to this work.

%
%________________________________________________________________
\bibliographystyle{plainnat}
\bibliography{R5_cits,references_prelaunch_forJI}

\end{document}